\documentclass[pra,twocolumn]{revtex4-1}
\usepackage{xcolor}
\usepackage{graphicx}
\usepackage{amssymb, amsmath, amsthm}
\usepackage{bbold}
\usepackage{dsfont}
\usepackage[percent]{overpic}

\usepackage[colorlinks]{hyperref}

\usepackage[authormarkup=none]{changes}
\definechangesauthor[name={a}, color=orange]{a} 
\definechangesauthor[name={c}, color=blue]{c} 

\def \diracspacing {0.7pt}
\newcommand{\bra}[1]{\langle #1 \hspace{\diracspacing} |} 
\newcommand{\ket}[1]{| \hspace{\diracspacing} #1 \rangle} 

\newcommand{\dg}{\dagger}
\newcommand{\cD}{\mathcal{D}}
\newcommand{\cH}{\mathcal{H}}
\newcommand{\cN}{\mathcal{N}}
\newcommand{\cL}{\mathcal{L}}

\newcommand{\rmd}{\mathrm{d}}

\begin{document}
	
\author{Alexandre Roulet}
\affiliation{Department of Physics, University of Basel, Klingelbergstrasse 82, CH-4056 Basel, Switzerland}

\author{Christoph Bruder}
\affiliation{Department of Physics, University of Basel, Klingelbergstrasse 82, CH-4056 Basel, Switzerland}

\title{Quantum synchronization and entanglement generation}
\begin{abstract}
	We study synchronization in a two-node network built out of the smallest possible self-sustained oscillator: a spin 1. We first demonstrate that phase locking between the quantum oscillators can be achieved, even for limit cycles that cannot be synchronized to an external semi-classical signal. Building upon the analytical description of the system, we then clarify the relation between quantum synchronization and the generation of entanglement. These findings establish the spin-based architecture as a promising platform for understanding synchronization in complex quantum networks.
\end{abstract}
\maketitle
\emph{Introduction.--}
Synchronization describes the tendency of self-sustained oscillators to adjust their intrinsic rhythm when weakly interacting~\cite{pikovsky01}. The first step towards studying synchronization in quantum networks is to address the case of two coupled limit cycles. Significant progress has been made in this direction by considering classically-inspired systems operating in the quantum regime, such as masers~\cite{armour16}, Van der Pol~\cite{tony13,walter14adp}, Kerr-anharmonic~\cite{niels17} or optomechanical oscillators~\cite{weiss16}. In this Letter, we depart from this approach and instead consider quantum systems with no classical analogue. Specifically, we consider a pair of spins 1, which has recently been shown to be the smallest possible unit for a quantum network~\cite{spin1}. An advantage of the spin-based architecture is the manageable size of the resulting Hilbert space, for which numerics are inexpensive in computational power and do not require any truncation. Moreover, synchronization being a perturbation effect, we can analytically derive all its features by performing a first-order expansion in the interaction strength, yielding a simple description of the phenomenon.

Previous studies~\cite{tony13,lee14} found that the phase locking of two Van der Pol oscillators coupled via a coherent interaction does not survive in the quantum regime. We thus first address the question whether the spins are able to phase lock at all. Exploring the different parameter regimes, we find that limit cycles with reversed dissipation rates synchronize best, while identical units are unable to adjust each other. More remarkably, phase locking is also achieved for limit cycles which could not be synchronized to an external semi-classical signal~\cite{spin1}. 

In the presence of noise, the tendency towards phase locking of
oscillators can be assessed by computing the distribution of their
relative phase~\cite{pikovsky01}, which for quantum systems is
contained in the phase-space quasidistribution. This method is unambiguous as it is explicitly formulated with respect to the phase variable of interest~\footnote{We note that alternative methods can be found in the literature, such as the indirect probe of Ref.~\cite{hush15} or the semiclassical quantity introduced in~\cite{mari13}, at the expense of loosing the one-to-one correspondence between the measure and the presence of synchronization.}. Yet, it would be appealing to identify a parameter which is not tied to a specific system, while still providing a conclusive measure of phase locking. The fundamental question at stake behind this objective is to grasp the essence of the synchronization phenomenon when applied to quantum systems, thereby revealing the common features shared by all types of networks in this regime. Observing that at the core of phase locking lies the creation of correlations between otherwise-independent oscillators, a measure of both classical and quantum correlations has been proposed as a universal meter applicable to any configuration: quantum mutual information~\cite{ameri15}. In parallel, it has been suggested that the presence of entanglement between the units certifies the onset of synchronization, thereby yielding the equivalent of the classic Arnold tongue based on purely information-theoretic reasoning~\cite{lee14}. Such a correspondence is reminiscent of the proposed use of entanglement as a quantum signature of chaos, recently explored experimentally in~\cite{chaudhury09,neill16}. Entanglement has also been suggested as a necessary ingredient in conformal field theories for the emergence of space-time in the corresponding anti-de Sitter space which describes gravity~\cite{vanRaamsdonk10,cowen2015}.

In this work, we take advantage of the simplicity of the system at hand and derive analytically the relation between synchronization and entanglement. We find that while both the quantum mutual information and the presence of entanglement appear to reproduce the Arnold tongue, they do not provide a conclusive measure of synchronization. An explanation in terms of the Schmidt decomposition is provided and an illustration of this result is shown where the entanglement is found to increase as synchronization is suppressed.

\emph{The model.--} 
\begin{figure}
	\includegraphics[width=0.9\columnwidth]{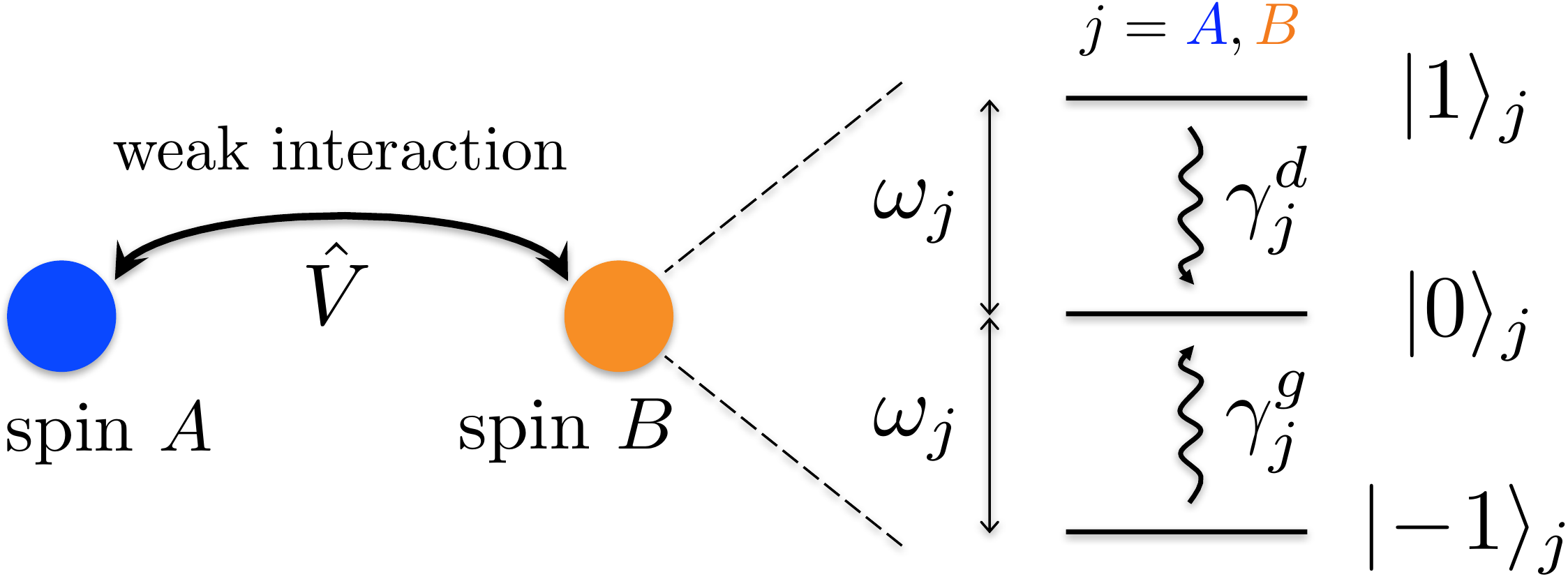}
\caption{\label{fig:2spins} Synchronization of a pair of
  self-sustained oscillators. Each spin 1 is continuously stabilized
  to its respective limit cycle. The question is whether the
  oscillators will tend towards locking their relative phase when
  weakly coupled despite not being necessarily identical.}
\end{figure}
As illustrated in Fig.~\ref{fig:2spins}, we consider two spins 1 weakly interacting via the exchange Hamiltonian
\begin{equation}\label{eq:v}
	\hat{V}=i\hbar\frac{\varepsilon}{2}\left(\hat{S}_A^+\hat{S}_B^- - \hat{S}_B^+\hat{S}_A^-\right) ,
\end{equation}
where $\hat{S}_j^+$ ($\hat{S}_j^-$) is the raising (lowering) operator of spin $j$. This corresponds to replacing the semi-classical drive of an externally-driven unit by a second quantum unit. At the same time, each spin is independently stabilized to their respective limit cycle by the Lindblad superoperator	$\cL_j\hat{\rho}=\frac{\gamma_j^g}{2}\cD[\hat{S}_j^+\hat{S}_j^z ]\hat{\rho}+ \frac{\gamma_j^d}{2}\cD[\hat{S}_j^-\hat{S}_j^z ]\hat{\rho} $
with $\cD[\hat{\mathcal{O}}]\hat{\rho}=\hat{\mathcal{O}}\hat{\rho} \hat{\mathcal{O}}^\dg-\frac{1}{2}\left\{\hat{\mathcal{O}}^\dg\hat{\mathcal{O}},\hat{\rho}\right\}$, and the gain $\gamma_j^g$ and damping $\gamma_j^d$ rates~\cite{spin1}. Consequently, the evolution of the system is described by the master equation
\begin{equation}\label{eq:master}
	\dot{\hat{\rho}}= -\frac{i}{\hbar}\left[\hat{V},\hat{\rho}\right]+\sum_{j=A,B}-i\left[\omega_j\hat{S}_j^z,\hat{\rho}\right]+\cL_j\hat{\rho} ,
\end{equation}
where the precession frequencies $\omega_j$ could in principle be
mismatched, $\Delta=\omega_A-\omega_B \ne 0$.

In the absence of interaction, the steady-state description of the
system is given by $\hat{\rho}_0=\ket{0}\bra{0}_A\otimes
\ket{0}\bra{0}_B$, with both spins appearing to possess identical
limit cycles. However, it is important to note that the spins, while
being stabilized to the same target state, do not react in the same
way to perturbations. In practice, the way they return to their limit
cycle is determined by the relative strength of gain $\gamma^g_j$ and damping $\gamma^d_j$. As we will see below, the possibility of stabilizing the spins differently will crucially impact the achievable phase locking, and also be a key component in understanding the role of entanglement.

Adopting the techniques introduced in Ref.~\cite{spin1}, we use the Husimi Q function as a phase-space distribution from which we define the following measure of synchronization
\begin{align}\label{eq:syncMeas}
	&S_\text{rel}(\phi)= -1/2\pi +\\ &\int_0^{2\pi}\!\!\!\rmd\phi_B\!\int_0^\pi\!\!\!\rmd\theta_A\!\int_0^\pi\!\!\!\rmd\theta_B\, \sin\theta_A\sin\theta_B\, Q(\theta_A,\theta_B,\phi+\phi_B,\phi_B) .\nonumber
\end{align}
This quantity is formulated explicitly in terms of the relative phase $\phi=\phi_A-\phi_B$ and allows to assess whether the spins show any tendency towards phase locking. If no fixed phase relation is established, e.g. in the absence of interaction, it remains identically null everywhere.

To provide a complete picture, we also consider the reduced state of each unit, which is obtained by performing a partial trace over the other spin $\hat{\rho}_A=Tr_B[\hat{\rho}]$ and similarly for $\hat{\rho}_B$. The corresponding phase distribution is then given by $p_j(\phi)=\int_0^\pi\!\rmd\theta\, \sin\theta\, Q_j(\theta,\phi)-1/2\pi$.

\begin{figure}
	\centering
	\begin{overpic}[width=0.48\columnwidth]{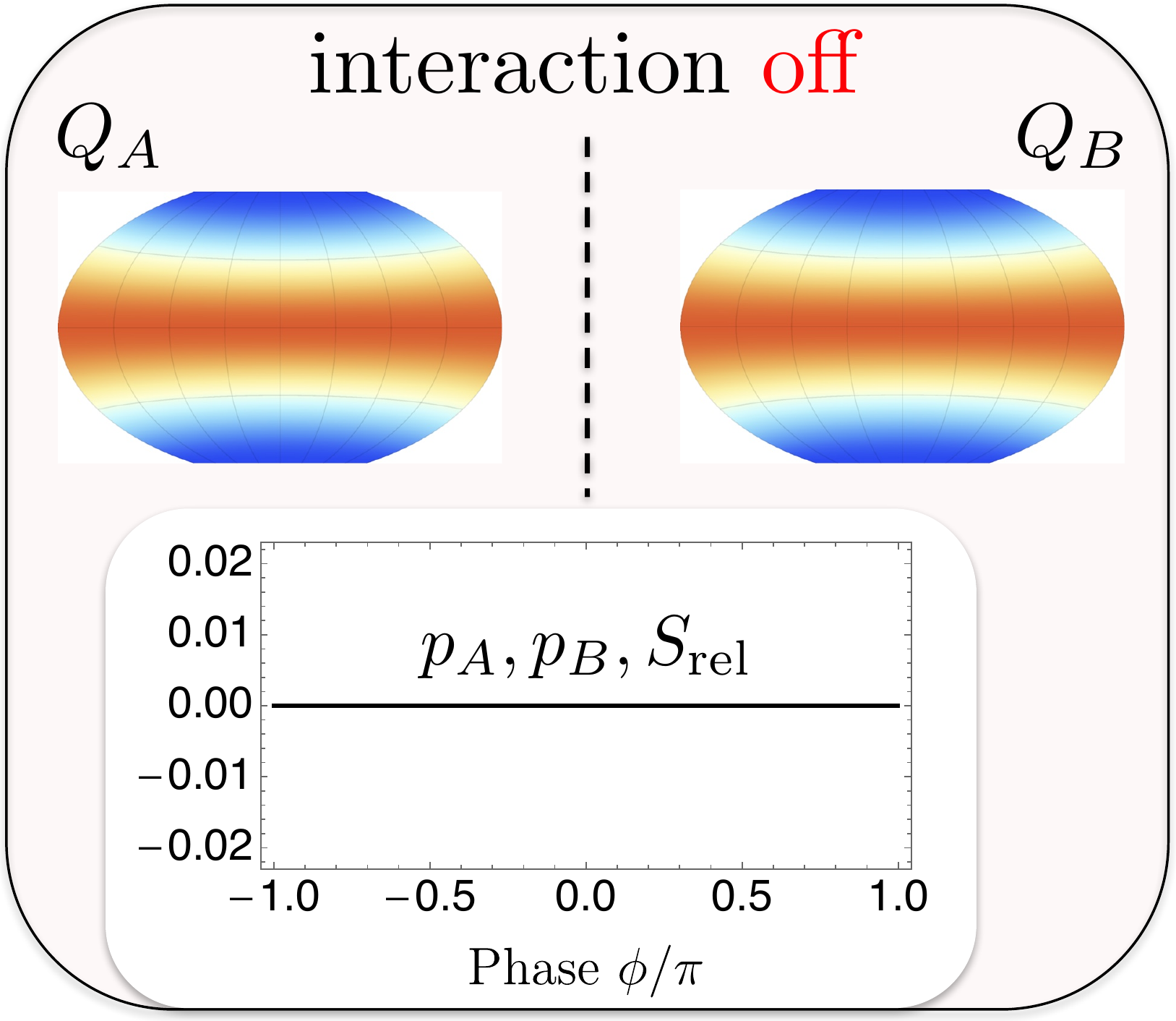}
		\put (0,90) {(a)}
	\end{overpic}\quad
	\begin{overpic}[width=0.48\columnwidth]{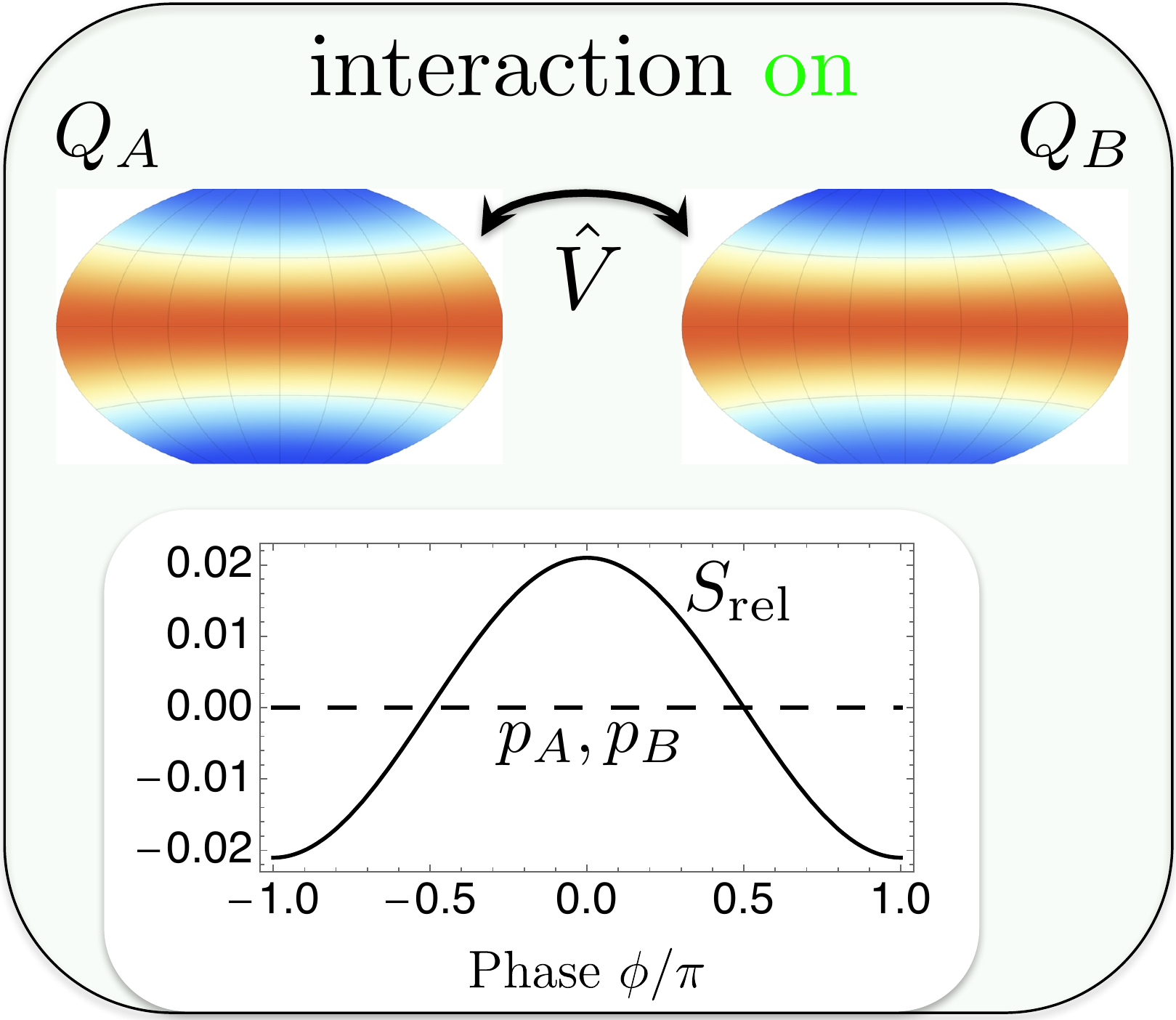}
		\put (0,90) {(b)}
	\end{overpic}
        \caption{\label{fig:arnoldS} Quantum synchronization for $\gamma^g_A/\gamma^d_A=\gamma^d_B/\gamma^g_B=\gamma^d_B/\gamma^d_A=100$.
(a) In the absence of interaction, \textit{i.e.},  $\varepsilon=0$,
both spins are stabilized to the equator without any phase preference
as shown by the sphere projection of their individual $Q$
functions. Consequently, their phases are uniformly distributed
($p_A=p_B=0$) and uncorrelated ($S_\text{rel}=0$). (b) Switching on
the interaction to $\varepsilon=0.1\gamma^d_A$ in the resonant case $\omega_A=\omega_B$, the distribution of the relative phase is peaked around $\phi=0$, corresponding to in-phase locking of the spins.}
\end{figure}
\emph{Quantum synchronization.--} 
The minimal dimension of the combined Hilbert space
$\text{dim}(\cH_A\otimes\cH_B)=9$ allows to solve numerically for the
steady state of the master equation~\eqref{eq:master} with no
effort. The results are illustrated in Fig.~\ref{fig:arnoldS} where
we have fixed the limit cycles to be inverted, \text{i.e.}, $\gamma^d_A=\gamma^g_B$ and $\gamma^g_A=\gamma^d_B$. Comparing Fig.~\ref{fig:arnoldS}(a) with (b), we find that upon switching on the interaction $\hat{V}$ the spins develop a pronounced in-phase locking of their relative phase. Moreover, each spin seems to be unaffected when tracing out over the other one, as can be seen from $Q_A$ and $Q_B$ which resemble the unsynchronized limit cycle with a symmetric distribution of the individual phases $p_A=p_B=0$. This follows from the reduced state of each spin being diagonal and therefore phase symmetric, a feature previously identified in the case of two coupled Van der Pol oscillators~\cite{walter14adp}. It can be understood in physical terms: as opposed to the synchronization of a single unit to an external coherent signal, the current system does not possess any phase reference. The spins therefore lock relatively to each other in a rotationally-invariant way. If one of the spins is traced out from the system, we consequently loose all the information on the relative phase and the remaining spin could be locked to any angle in a uniform way. We thus recover the initial limit cycle, which is not a coincidence but crucially follows from the requirement of not deforming it (moving away from the equator) when attempting to synchronize its phase~\cite{pikovsky01}.

\begin{figure}
	\centering
	\begin{overpic}[width=0.505\columnwidth]{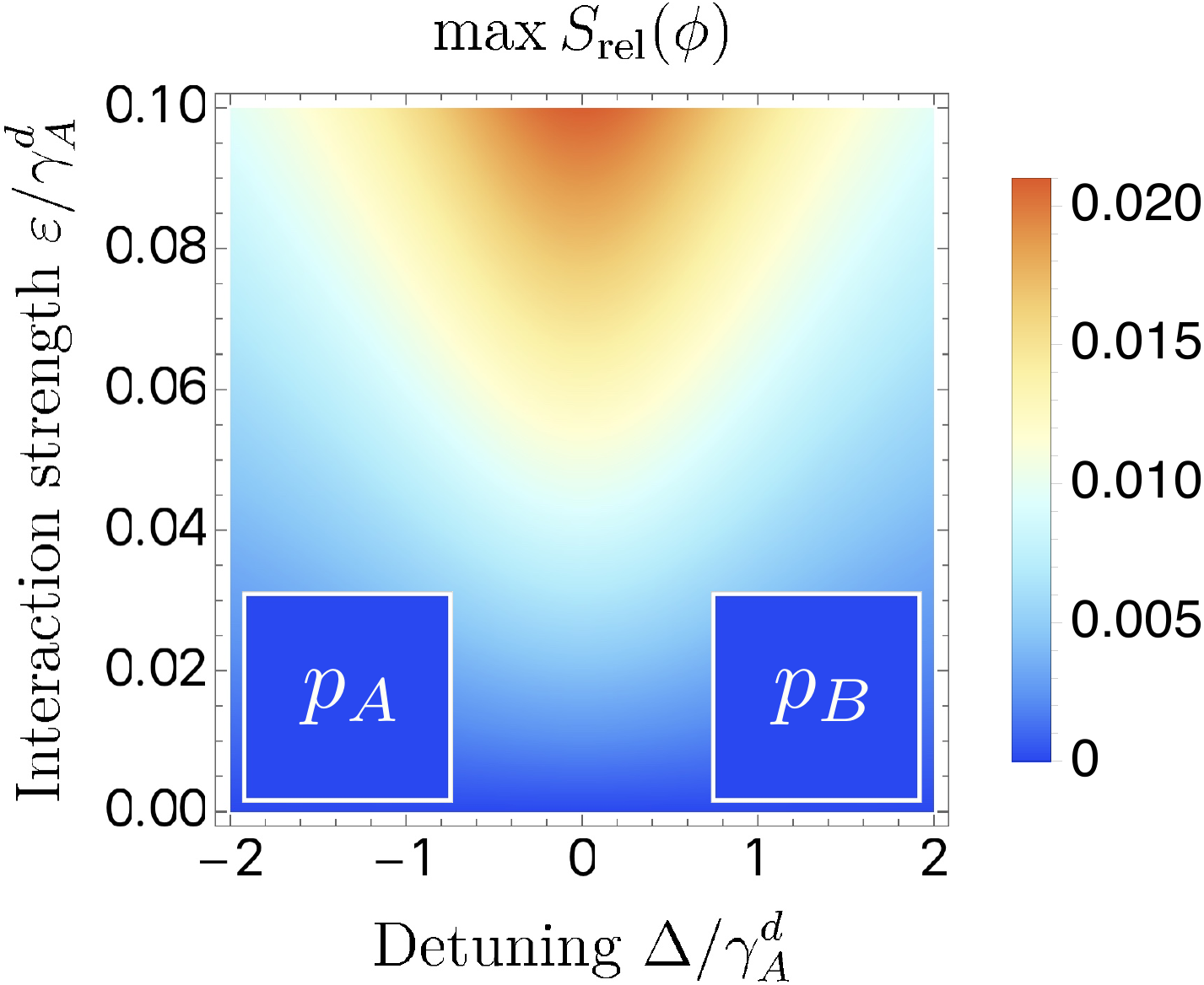}
		\put (0,84) {(a)}
	\end{overpic}\quad
	\begin{overpic}[width=0.455\columnwidth]{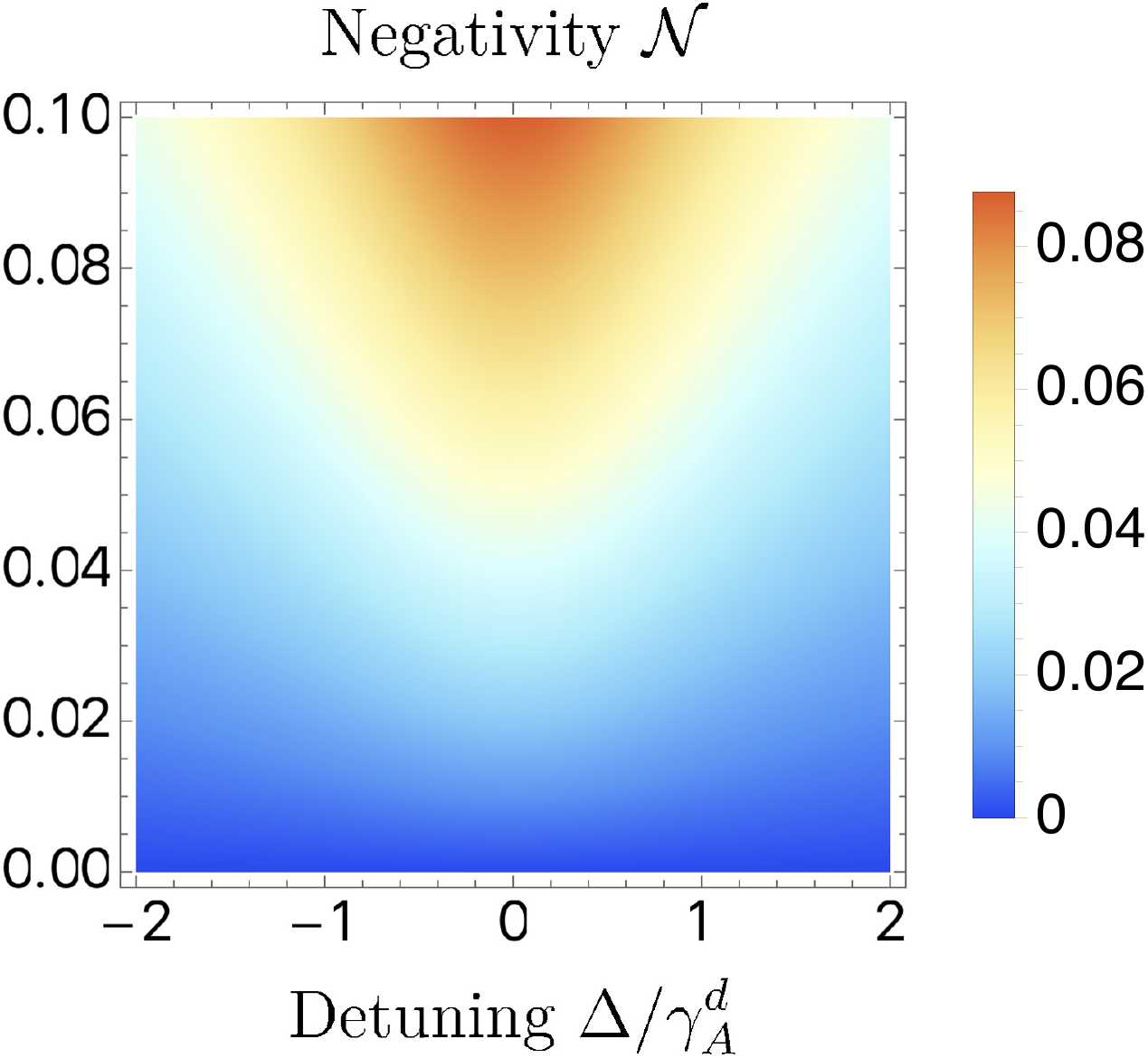}
		\put (-10,93) {(b)}
	\end{overpic}
\caption{\label{fig:arnoldIN} (a) The Arnold tongue, obtained by varying the interaction strength $\varepsilon$ and detuning $\Delta=\omega_A-\omega_B$, for the same parameters as in
  Fig.~\ref{fig:arnoldS}. The insets show that the individual phases remain uniformly distributed ($p_A=p_B=0$). (b) The negativity $\cN$ reproduces the Arnold tongue.}
\end{figure}

As expected from the general framework of synchronization, phase locking is found to survive upon detuning the frequency of the spins. Specifically, the stronger the interaction strength, the larger the range of detuning leading to significant localization of the relative phase, yielding the Arnold tongue shown in Fig.~\ref{fig:arnoldIN}(a). In fact, we find that the performance achieved for the present choice of parameters is comparable to that obtained in the case of a single spin subjected to an external signal~\cite{spin1}. It is well-known that for classical systems~\cite{pikovsky01}, these two scenarios are characterized by the same dynamical equation, describing in one case the relative phase between the spins, and in the other case the phase of the single externally-locked spin. To explore whether this holds in our quantum setup, we now turn to analytics.

Synchronization being a perturbation effect, its features can be captured by performing a first-order expansion in the interaction strength $\varepsilon$. We thus expand the state of the system around the initial uncorrelated limit cycles as $\hat{\rho}(t)\approx \hat{\rho}_0 + \varepsilon \hat{\mu}(t)$ with $\text{Tr}[\hat{\mu}]=0$. Discarding higher-order terms, the master equation~\eqref{eq:master} reads
\begin{equation}\label{eq:masterMu}
	\dot{\hat{\mu}}= \frac{1}{2}\left[\hat{S}_A^+\hat{S}_B^- - \hat{S}_B^+\hat{S}_A^-,\hat{\rho}_0\right]+\sum_{j=A,B}-i\left[\omega_j\hat{S}_j^z,\hat{\mu}\right]+\cL_j\hat{\mu} .
\end{equation}
The first term is a constant drive of the coherences $\mu_{\pm}=\bra{\pm1,\mp1}\hat{\mu}\ket{0,0}$ by the interaction Hamiltonian~\eqref{eq:v}. The second term rotates these coherences in the complex plane without coupling different matrix elements while the last term acts as a damping. We have thus reduced the problem of solving a master equation in a 9-dimensional Hilbert space to a set of two first-order differential equations for the non-trivial terms of the density matrix.

With this analytical description at hand, we can now derive the dynamics of the phase locking~\eqref{eq:syncMeas} as
\begin{align}
	\dot{S}_\text{rel}(\phi)&\approx\frac{9\pi\varepsilon}{128} \text{Re}\left[e^{i\phi}(\dot{\mu}_+ +\dot{\mu}_-^*)\right] \\
	&= \frac{9\pi\varepsilon}{128} \left(e^{-(\gamma^d_A+\gamma^g_B)t/2}-e^{-(\gamma^g_A+\gamma^d_B)t/2}\right)\cos(\phi-\Delta t) , \nonumber
\end{align}
where the approximate sign is a reminder that this result holds only
within the perturbation regime of interest. This is one of the main results of our Letter. As previously suspected
from the numerics, this expression is of the
same form as in the case of synchronization to an external signal~\cite{spin1}, up to a constant factor and the specific rates appearing in the exponentials. Yet, as we will show below, applying a semi-classical signal and coupling to a quantum system are not equivalent in the context of phase locking.

We focus here on the
resonant case $\Delta=0$ for which the coherences are real numbers
with $\mu_+ \ge 0 \ge \mu_-$. The synchronization measure is then proportional to their sum
\begin{align}\label{eq:sRes}
	S^{\Delta=0}_\text{rel}&(\phi)\approx\frac{9\pi\varepsilon}{128}(\mu_+ +\mu_-)\cos\phi \\
	&\xrightarrow[]{steady-state} \frac{9\pi\varepsilon}{128} \left(\frac{2}{\gamma^d_A+\gamma^g_B}-\frac{2}{\gamma^g_A+\gamma^d_B}\right)\cos\phi . \nonumber
\end{align}
Firstly, we find that the measure vanishes for spins with identical limit cycles $\gamma^d_A=\gamma^d_B$ and $\gamma^g_A=\gamma^g_B$. We note that this peculiarity has been previously noticed for the case of two Van der Pol oscillators brought to the quantum limit~\cite{tony13,lee14}, leading to the suspicion that phase locking between quantum units could only be obtained via dissipative coupling. Instead, we find that limit cycles with reversed dissipation rates are very-well able to lock, as illustrated in Fig.~\ref{fig:arnoldS} where $\gamma^g_A=\gamma^d_B\gg\gamma^d_A=\gamma^g_B$. 

Even more remarkably, Eq.\,\eqref{eq:sRes} demonstrates that a balanced limit cycle $\gamma^d_A=\gamma^g_A$, which was found to be unresponsive to an external signal~\cite{spin1}, phase locks equally well to an unbalanced one $\gamma^d_B\gg\gamma^g_B$ or $\gamma^d_B\ll\gamma^g_B$. We thus find that a second quantum spin 1 is able to synchronize a larger variety of limit cycles than if we were to replace it by a semi-classical coherent signal. It goes without saying that synchronization of a pair of oscillators is not a quantum effect. However, this result points to the fact that when considering a quantum unit, one can improve the phase locking by taking full advantage of the quantum framework governing the system instead of restricting parts of the setup to be classical~\cite{kwek18}.

\emph{Entanglement generation.--} 
When discussing Fig.~\ref{fig:arnoldS}(b), we touched on the fact that the information about synchronization is contained in the state of the total system and is lost when restricting the available knowledge to the reduced states of the individual units. This hints to a possible relation between the onset of synchronization and the generation of entanglement between the spins. On this basis, the natural question that arises is whether one can use the presence of such correlations as an indicator of quantum limit cycles having synchronized~\cite{lee14}. To test this hypothesis, we employ the negativity $\cN = (||\hat{\rho}^{T_A}||_1-1)/2$ as a measure of entanglement for our bipartite system, where $||\hat{\rho}^{T_A}||_1$ is the trace norm of the
partially-transposed state~\cite{vidal02}. We note that there exists a zoo of
measures for bipartite mixed states which are not always equivalent
and often hard to compute~\cite{plenio07}. There is no ambiguity in the present context where both numerics and first-order analytics will indicate that the state is virtually pure. Following the suggestion of Ref.~\cite{ameri15}, we also consider the quantum mutual information, a measure of all types of correlations between the units given by $I=S(\hat{\rho}_A)+S(\hat{\rho}_B)-S(\hat{\rho})$
where $S(\hat{\mathcal{O}})$ stands for the von Neumann entropy.

As a first sanity check, we investigate whether these
purely information-theoretic measures are able to reproduce the Arnold
tongue. This is shown in
Fig.~\ref{fig:arnoldIN}, where the negativity is found to follow
the desired pattern. The quantum mutual information $I$ behaves similarly. To confirm this visual impression, we perform a
linear regression using the least-square method over the $101\times 101$ data points composing the plots of Fig.~\ref{fig:arnoldIN}. This yields $\max S_\text{rel}(\phi)\approx 0.23 \cN$ with $R^2\approx1$, where a coefficient of determination $R^2$ close to unity validates the linear relation. The entanglement measure $\cN$ thus appears to have a perfect predictive power on the
degree to which the spins are phase locked. Likewise, the correlations captured by the quantum mutual information $I$ provide a very accurate measure $\max S_\text{rel}(\phi)\approx 0.27 I$ with $R^2\approx0.96$. Importantly, the absence of an offset in both cases indicates a one-to-one correspondence.

To resolve whether this constitutes a valid measure, we turn once more to the first-order expansion, aiming this time at establishing an analytical relation between synchronization and entanglement. From our solution to the recast form of the master equation~\eqref{eq:masterMu} which depends only on the coherences $\mu_\pm$, the state of the system is readily found to be approximately in the pure entangled state $\hat\rho\approx \ket{\Psi}\bra{\Psi}$ where
\begin{equation}\label{eq:psi}
	\ket{\Psi}\propto\ket{0,0}+\varepsilon\mu_+\ket{1,-1}+\varepsilon\mu_-\ket{\!-\!1,1} .
\end{equation}
The absolute value of the coherences $|\mu_\pm|$ are therefore directly related to the Schmidt coefficients of the pure state decomposition, yielding $\cN\approx\varepsilon\big(|\mu_+|+|\mu_-|\big)$.

\begin{figure}
	\centering
	\begin{overpic}[width=0.505\columnwidth]{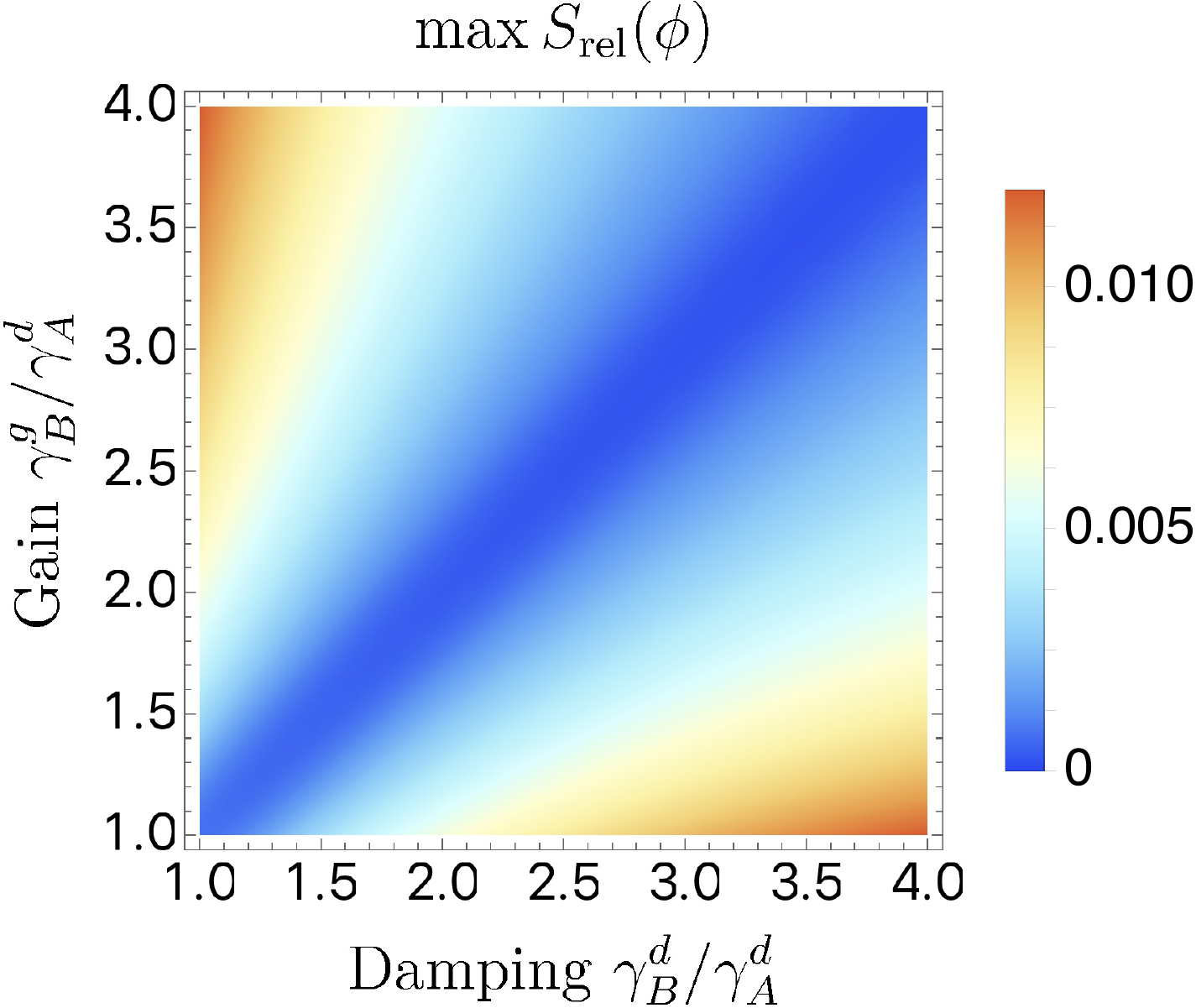}
		\put (0,85) {(a)}
	\end{overpic}\quad
	\begin{overpic}[width=0.455\columnwidth]{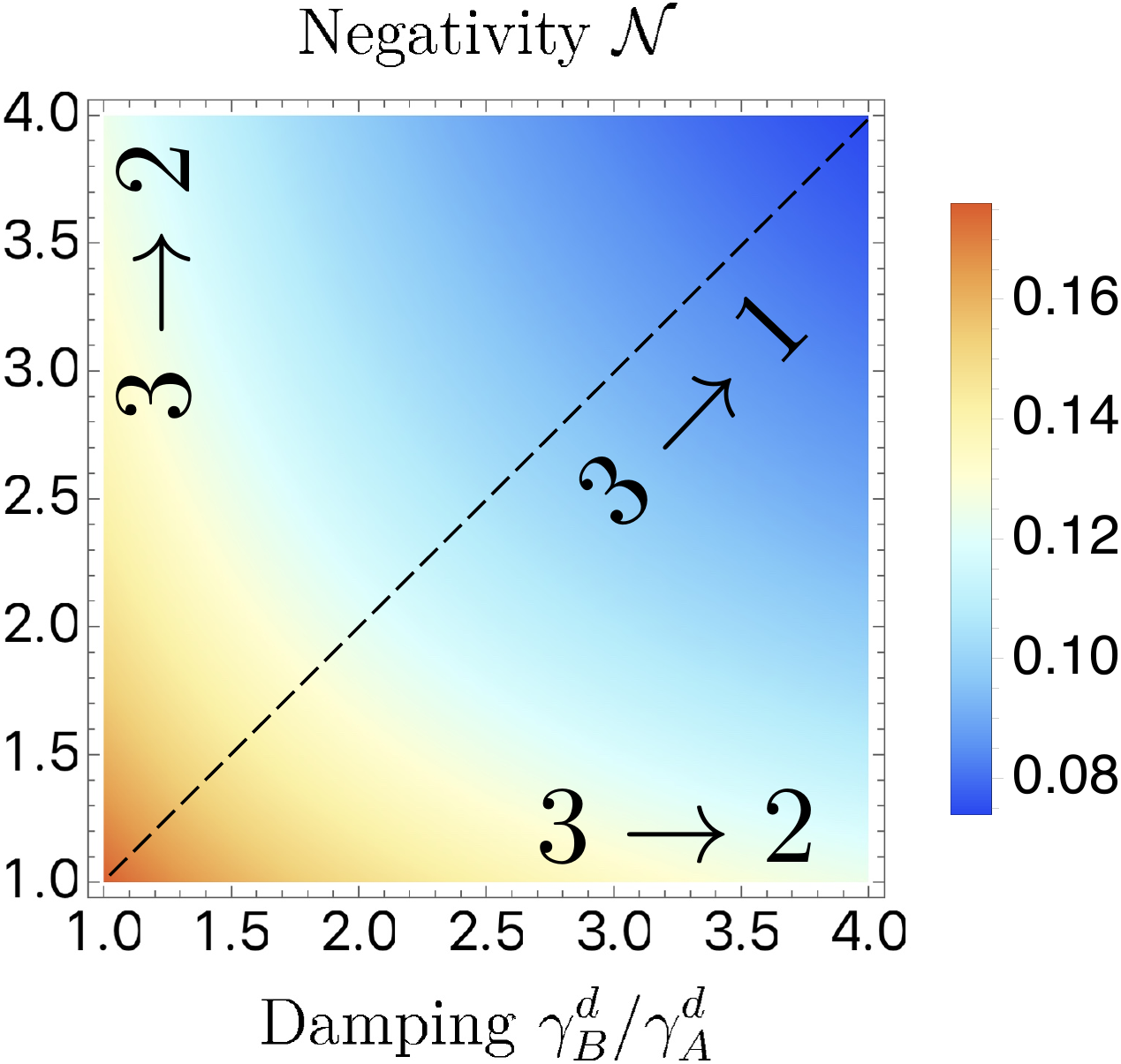}
		\put (-10,94.5) {(b)}
	\end{overpic}\vspace{0.2cm}
	\begin{overpic}[width=0.8\columnwidth]{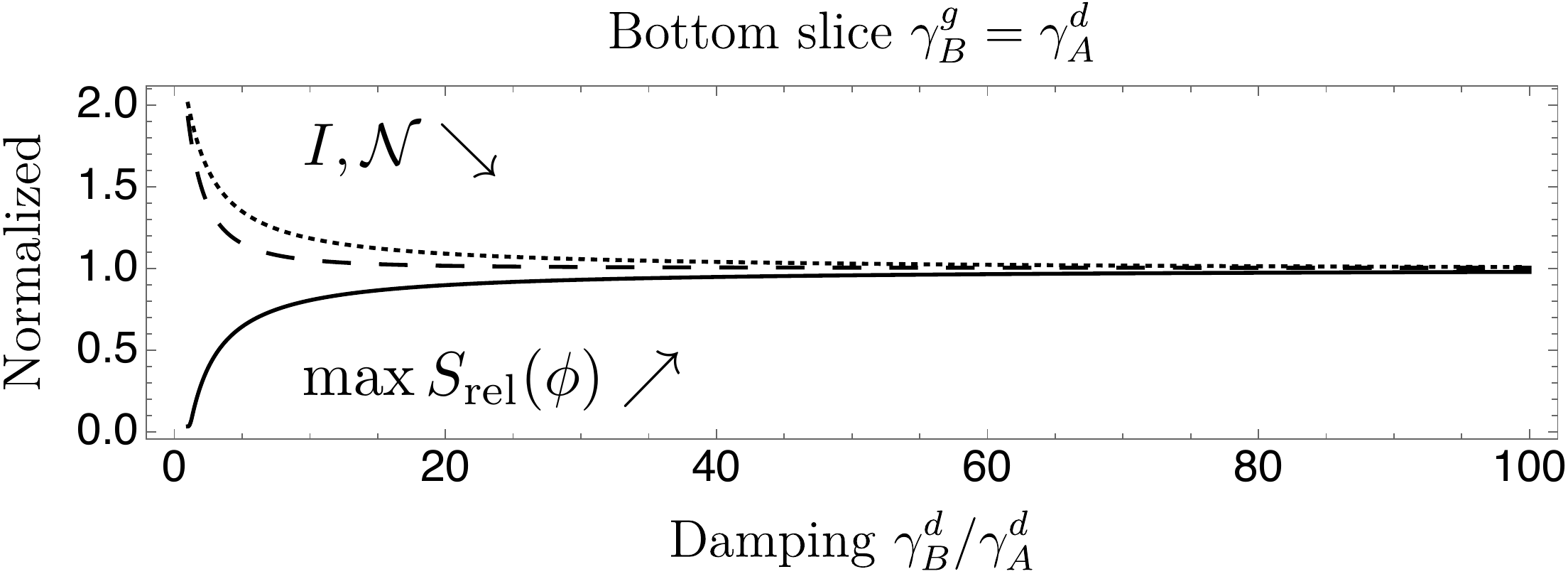}
		\put (-10,30) {(c)}
	\end{overpic}
\caption{\label{fig:balanced} (a) Synchronization compared with (b) negativity $\cN$ on resonance $\Delta=0$ for $\varepsilon=0.1\gamma^d_A$ and $\gamma^g_A=\gamma^d_A$. The numbers in (b) indicate the Schmidt rank of the state given in Eq.\,\eqref{eq:psi}. The quantum mutual information $I$ follows the same pattern as the negativity. No phase locking would be achieved in this regime by coupling spin $A$ to a semi-classical coherent signal. (c) Focusing on an extended cut of the density plots, both $I$ (dashed) and $\cN$ (dotted) are reduced by half when phase locking is established between the spins. The quantities are normalized with respect to the values obtained for locked reversed limit cycles (middle top point of Fig.\,\ref{fig:arnoldIN}).}
\end{figure}

This result demonstrates that
synchronization between the spins cannot be established without the generation of entanglement. But is the converse also true? Namely, is witnessing the presence of entanglement providing a conclusive measure of synchronization? Figure~\ref{fig:arnoldIN} points towards a positive answer. In fact, the value of $0.23$ obtained numerically from the linear regression on the negativity is nothing else than the coefficient $9\pi/128\approx 0.22$ entering the synchronization measure to first order. Yet, focusing on the resonant case to compare with Eq.\,\eqref{eq:sRes}, we find that the negativity is proportional to the difference between the coherences instead of their sum
\begin{equation}
	\cN^{\Delta=0}\approx\varepsilon\left(\mu_+-\mu_-\right) \neq \frac{\max S^{\Delta=0}_\text{rel}(\phi)}{9\pi/128} .
\end{equation}
Note that there is no contradiction so far as $|\mu_+|\gg|\mu_-|$ for the reversed limit cycles considered in Fig.\,\ref{fig:arnoldIN}. However, this result indicates that there exist regimes where the spins can get significantly entangled without being able to lock their relative phase.

This is illustrated in Fig.~\ref{fig:balanced}, where the limit cycle of the first spin is fixed to be balanced, $\gamma^d_A=\gamma^g_A$. The information-theoretic measures manifestly differ from the expected behavior. Starting with the second spin to be balanced as well, we find no phase locking as the coherences are exactly opposed $\mu_+=-\mu_-$. On the other hand, the spins are significantly entangled with a corresponding Schmidt rank of 3. As we vary the rates to unbalance the second limit cycle, synchronization gets established as previously discussed. But at the same time the entanglement is found to decrease, which is due to one of the coherences being significantly reduced, and the Schmidt rank consequently dropping to 2. Eventually, all measures reach the same values as in the case of locked reversed limit cycles. This shows that the presence of entanglement and a fortiori of correlations between the spins is not exclusively related to the onset of synchronization. Reversing the premise, we note however that synchronization is a valid witness of the entanglement generated by the weak interaction~\eqref{eq:v}, identifying certain states of Schmidt rank 2 while being blind to states of Schmidt rank 3.

\emph{Conclusion.--} 
We have demonstrated that two quantum units are able to lock their
phase under a weak Hamiltonian interaction. The spin-based architecture therefore paves the way for the investigation of synchronization in large quantum networks with minimal unit size. In this context, the possibility of frustrating neighboring sites by adjusting the way their limit cycles are stabilized to the very same target state promises a rich collective behavior. 

Our analytical description of the system also provides access to the
role played by the entanglement generated between the spins. In
particular, we have shown that while synchronization can be used to
certify the presence of entanglement in the system, the converse is
not necessarily true, thereby ruling out the possibility of using
simple measures of correlations as a generic order parameter for
quantum networks. Yet, there is the exciting possibility of using the presence of phase locking as an entanglement witness for larger networks, certifying pair-wise or perhaps even multipartite entanglement. Moreover, the present work provides an elementary setup to test candidates for a refined information-theoretic measure of synchronization.
\begin{acknowledgments}
We would like to thank J.~Kaniewski for insightful discussions on measures of entanglement.
This work was financially supported by the Swiss SNF and the NCCR Quantum Science and Technology.
\end{acknowledgments}

\bibliography{2spin1Bib}

\end{document}